\documentclass[aps,prb,superscriptaddress,reprint]{revtex4-1}
\usepackage{graphicx}
\usepackage{amsmath}
\usepackage{amssymb}
\usepackage{amsfonts}
\usepackage{filecontents}
\usepackage{color}
\usepackage[normalem]{ulem}
\usepackage{soul}
\usepackage{epstopdf}
\usepackage{hyperref}
\usepackage{fancyhdr}
\definecolor{gold}{rgb}{0.85,.66,0}

\begin{document}

\title{Phonon sidebands of color centers in hexagonal boron nitride}

\author{P. Khatri}
\affiliation{College of Engineering, Mathematics and Physical Sciences,University of Exeter, Exeter EX4 4QF, United Kingdom}

\author{I. J. Luxmoore}
\affiliation{College of Engineering, Mathematics and Physical Sciences,University of Exeter, Exeter EX4 4QF, United Kingdom}

\author{A. J. Ramsay}
\affiliation{Hitachi Cambridge Laboratory, Hitachi Europe Ltd., Cambridge CB3 0HE, United Kingdom}

\date{\today}

\begin{abstract}{
Low temperature photoluminescence spectra of a color center in hexagonal boron nitride are analyzed. The acoustic phonon sideband can be described by a deformation coupling proportional to strain   to a phonon bath that is effectively two dimensional. The optical phonon band is described by Fr\"{o}hlich coupling to the LO-branches, and a deformation coupling proportional to lattice displacement for the TO-branch. The resonances expressed in the optical band vary from defect to defect, in some emitters, coupling to out-of-plane polarized phonons is reported.   }
\end{abstract}

\maketitle





\section{Introduction}

The photon indistinguishability of a  single-photon source based on a quantum dot or color center is degraded by phonon-assisted emission, which may be rejected at the expense of brightness using a wavelength filter, or improved through cavity enhancement of the zero-phonon line \cite{Iles_Smith_nphoton}.  The electron-phonon interactions also determine the fidelity with which the emitter can be prepared in the excited state \cite{Ramsay_prl,Quilter_prl}. In this respect, due to their large carrier-wavefunctions, InAs quantum dots have the best optical coherence properties with lifetime limited dephasing, and about 96\% emission into the zero-phonon line \cite{Borri_prl,Borri_prb}. However, due to the large carrier-wavefunctions, the cut-off energy of the electron-phonon interaction $\hbar\omega_c=1.3 ~\mathrm{meV}$ \cite{Ramsay_prl2} restricts this to low temperatures, since $\hbar\omega_c/k_B=16~\mathrm{K}$.

Recently, high temperature single-photon emission of color-centers in hexagonal boron nitride (h-BN) have been reported \cite{Tran_nnano,Martinez_prb} up to $800^{\circ}C$ \cite{Kianinia_acsph}. Whilst there are reports of room temperature single photon emission in InGaN quantum dots \cite{Holmes_nano,Holmes_acsphot}, color centers in diamond \cite{Mizuochi_nphoton}
 and SiC \cite{Lohrmann_ncomm}, h-BN stands out for the high ZPL fraction of the emission. h-BN is a layered material with a graphene-like lattice. It has a large bandgap, and is used as an insulator in 2D electronics. This raises the question of whether the highly anisotropic crystal structure leads to a phonon bath that is effectively 3D or 2D, and if so, does the reduced dimensionality have an advantage for the optical coherence properties of the emitter. The effects of dimensionality on acoustic-phonon assisted emission of quantum dots in nanostructures has recently been investigated theoretically \cite{Tighineanu_prl,Palma_prsoc}. The issue has been experimentally investigated for quantum dots in Carbon nanotubes with a 1D-phonon bath \cite{Galland_prl}. Recently, Vuong {\it et al} \cite{Vuong_prl} have analyzed the acoustic phonon sidebands of UV-emitting defect-bound excitons in bulk h-BN, and explain the results in terms of a 3D acoustic phonon-bath with an angle-averaged speed of sound. There are a couple of reports that fit the acoustic and optical sidebands at room \cite{Exharos} and low temperature \cite{Feldman_ArXiv}.

Here we analyze the phonon sidebands of emission from color center in multi-layer flakes of h-BN in detail. Away from the ZPL, the acoustic phonon sideband can be described by an in-plane deformation coupling to phonon-bath that is effectively two-dimensional due to a highly anisotropic acoustic-phonon dispersion that is largely independent of out-of-plane momentum. At low temperature, the optical phonons give rise to multiple resonances at points where the phonon group-velocity is low. The resonances expressed vary from emitter to emitter, and may provide clues on the shape of the defect wavefunctions involved.

\section{Photoluminescence measurements}

The sample consists of few-layer flakes of hBN (Graphene Supermarket) drop-cast onto a Silicon substrate, which then undergoes a rapid thermal anneal in nitrogen atmosphere, ramping up to $850^{\circ} C$ in 7 minutes and holding for 8 minutes, before being allowed to cool. The studied samples are most likely multi-layer flakes of a few $\mathrm{\mu m^2}$ area. For a 532 nm pump laser, the density of emitters is low, under 1 per 100 $\mathrm{\mu m^2}$. We focus on two color centers emitting close to 2.17 eV.

\begin{figure}[ht!]
\begin{center}
\vspace{0.2 cm}
\includegraphics[width=1\columnwidth]{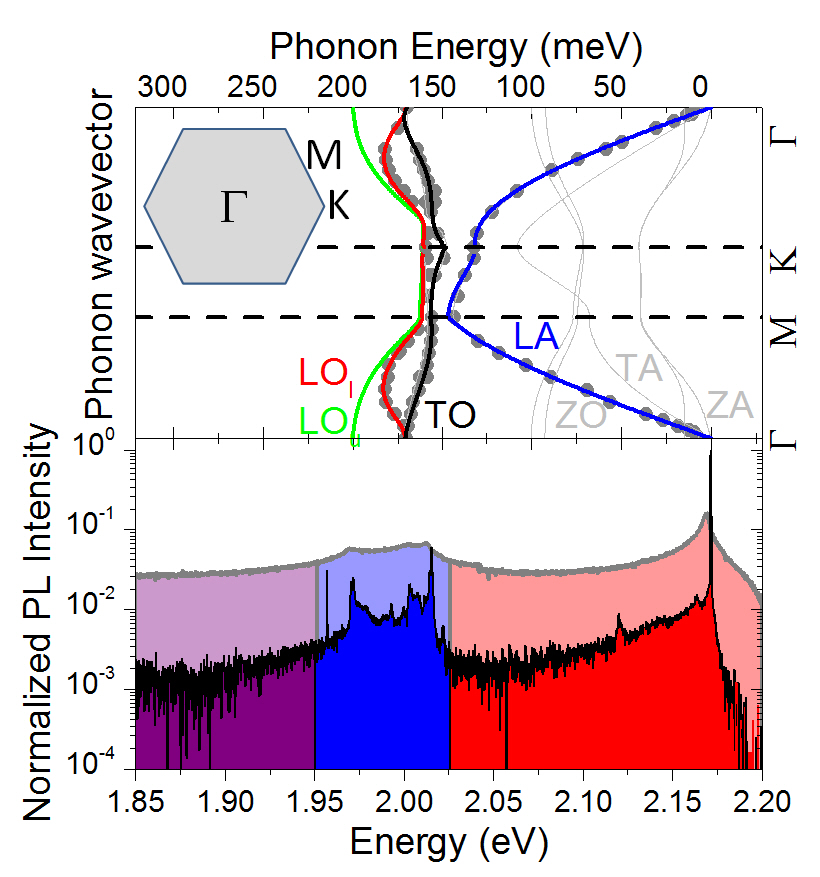}
\vspace{0.2 cm}
\end{center}
\caption{(a) Comparison of PL at 20 K and 275 K for emitter I. The ZPL can be identified by its large drop in intensity with temperature. (above) By comparison to the phonon dispersion curves \cite{Serrano_prl} the acoustic and optical sidebands can be identified. (inset) 2D Brillouin zone. }\label{fig:fig1}
\end{figure}

Fig. 1(a) compares the micro-photoluminescence spectra of emitter I using a 532nm pump at temperatures of 20  and 275 K. Close to room temperature, the spectrum consists of a bright, narrow line at 2.171 eV. We attribute this peak to the zero phonon line (ZPL) emission of the color centre, since the intensity of the peak is sensitive to temperature. The energy of the ZPL is 159 meV less than the photon energy of the pump laser, similar to the energy of an optical phonon. Defects emitting at similar energies have been previously reported \cite{Jungwirth_nl}. Although the setup is polarization dependent,  both the emission and absorption are found to be approximately co-linearly polarized, consistent with previous reports \cite{Jungwirth_prl}. At 275 K, there is a slight asymmetry of the ZPL, which is clearly revealed as the temperature is decreased. In fig. 1(a), the spectra is color-coded to indicate  red-detuned sidebands due to acoustic phonon emission assisted radiative recombination (detuning $<$150 meV)(red), and optical phonon emission assisted radiative recombination (150$<$detuning$ <$200 meV)(blue).

At  low temperature, approximately 18\% of the emitted photons are from the zero phonon line at 571 nm.  This is high compared to (12\%, 3.7\% at 575 nm, and 682 nm, ) reported in ref. \cite{Jungwirth_nl}, and considerably smaller than the (80\% at 623nm)reported in ref. \cite{Tran_nnano}.   The coupling of the color centre to the vibronic modes, as evident in the temperature dependent spectra of Fig. 1, leads to the observed reduction in the relative intensity of the ZPL as the temperature is increased. About 19 \% of the emission is into the optical phonon band, indicating that coupling to optical phonons is efficient, and the broad acoustic phonon band accounts for the rest.

\section{Analysis of acoustic phonon sideband}

To analyze the phonon sidebands we compare the data to a model that considers a two-level system coupled to a bath of phonons. If the system is excited into the upper energy level at time zero, the electronic polarization, acting as the light source  decays as:
\begin{equation}
P(t)=~exp(-\gamma(t)-i\Phi(t)),
\end{equation}
where \cite{Vagov_prb}
\begin{eqnarray}
\gamma(t)=\frac{\gamma_{ZPL}t}{2}+ \nonumber \\ \sum_{a}\int \frac{V_dd^dk}{(2\pi)^d} \vert\frac{g_{k,a}}{\omega_k}\vert^2(1-\cos{\omega_{k,a}t})(2N(\omega_{k,a},T)+1), \nonumber \\
\Phi(t)=\sum_{a} \int \frac{V_dd^dk}{(2\pi)^d} \vert\frac{g_{k,a}}{\omega_k}\vert^2\sin{\omega_{k,a}t}. \nonumber
\end{eqnarray}
$g_{k,a}$ gives the electron-phonon coupling with a phonon of wave-vector $k$ in band $a$, with energy $\hbar\omega_{k,a}$. $N(\omega,T)$ is the Bose-Einstein distribution. The PL spectrum is then calculated as $S(\omega)=\mathrm{real}(\int_0^{\infty}dt e^{-i\omega t}P(t))$. The sum over $k$ is expressed as an integral over a d-dimensional k-space of volume $V_d$. The energy dispersion is found from a fit to data in ref. \cite{Serrano_prl}. To interpolate between the $\Gamma-K$ and $\Gamma-M$ directions we approximate the dispersion in terms of the magnitude $k$ and direction $\theta$ of the k-vector as:
\begin{eqnarray}
\omega^2(k,\theta)\approx
\frac{1}{2}(\omega^2_{\Gamma-K}(k)+\omega^2_{\Gamma-M}(k))+ \nonumber \\
\frac{1}{2}(\omega^2_{\Gamma-K}(k)-\omega^2_{\Gamma-M}(k))\cos{6\theta}.
\end{eqnarray}



To describe the electron-phonon coupling, we expand the electron-phonon coupling in terms of wave-vector $k$, and consider the first two-terms.
\begin{eqnarray}
g_{k,a}=\frac{1}{\sqrt{2\rho_dV_d\hbar\omega_{k,a}}}(i.\mathcal{M} +k\mathcal{D})f(k)
\end{eqnarray}
The pre-factor normalizes the energy of the phonon-mode to $\hbar\omega$. $\rho_d$ is the d-dimensional mass-density. $f(k)$ is a form-factor and is the overlap of the electron density and the phonon wavefunction.
The $\mathcal{M}$-term describes a piezo-like change in the transition energy proportional to the lattice displacement, and the $\mathcal{D}$-term a deformation-like coupling proportional to the strain. For acoustic phonons, the $\mathcal{D}$-term dominates since the strain changes the separation between neighboring lattice sites. In the case of a quantum dot, these are usually properties of the quantum dot host material \cite{Krummheuer_prb}, but here are likely to be properties of the defect.

To help identify the dimension and coupling mechanism of the acoustic phonons, we consider the low k regime, where the dispersion is linear. In this case the coupling term can be expressed as:
\begin{equation}
\int \frac{V_dd^dk\vert g_k\vert^2}{(2\pi)^d}\rightarrow \alpha_n \int dk k^n g(k)
\end{equation}
where $n=d-1+2c-1+4p$ describes the `order' of the coupling.  For piezo and deformation-like coupling, $c=0,1$, respectively.  For an upper and lower energy-level with different electron-phonon coupling constants, $p=0$,$f(0)=1$ and if the coupling-constants are the same, $p=0$,$f(0)=0$. The `order' determines the qualitative shape of the phonon sideband, $n=1$ is often referred to as the `Ohmic' case \cite{Palma_prsoc}.

\begin{figure}[ht!]
\begin{center}
\vspace{0.2 cm}
\includegraphics[width=1\columnwidth]{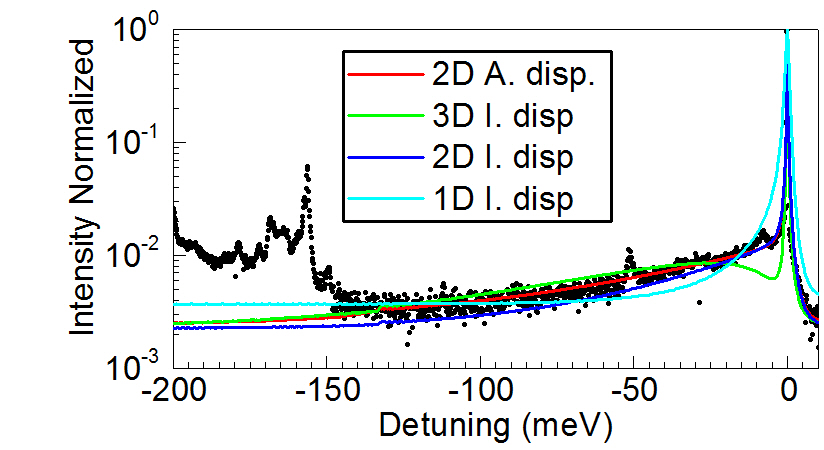}
\vspace{0.2 cm}
\end{center}
\caption{ Close up of acoustic phonon sideband. Note that sideband decays exponentially with detuning. Lines are calculations that assume electron phonon interactions of order n=1 (cyan),2 (blue),3 (green). Qualitatively, the n=2 case fits best. The red line shows the case n=2, using the full anisotropic dispersion curves, and gives a slightly better fit to data. The parameters used are: $D_{1D}=4 ~\mathrm{eV}, \rho_{1D}=1.65\times 10^{-16}~\mathrm{kg.m^{-1}}, \sigma_{1D}=0.21~\mathrm{nm}$; $D_{2D}=10.5~\mathrm{eV}, \rho_{2D}=0.76~\mathrm{mg.m^{-2}},\sigma_{2D}=0.35~\mathrm{nm}$; $ D_{3D}=26 ~\mathrm{eV}, \rho_{3D}=2.18\times 10^3~\mathrm{kg.m^{-3}}, \sigma_{3D}=0.49~\mathrm{nm}$.  }\label{fig:fig2}
\end{figure}

In fig. 2(a), a close-up of the acoustic phonon-sideband is shown. The exponential decrease of the phonon-sideband with red-detuning implies an exponential form-factor, $f^2(k)=e^{-k\sigma}$. The absence of resonances near 75 meV and 100 meV, implies that the ZO-phonons do not play a significant role. To identify the electron-phonon interaction, a set of calculations made for different power laws for $n=1,2,3$, are shown alongside the data in fig. 2(a). Only $n=2$ can describe the data. Therefore the acoustic sideband arises from deformation coupling to a phonon bath that is effectively two-dimensional. Since the form-factor restricts the contributing phonons to low-k, it is difficult to distinguish between contributions from TA and LA phonons.

An effective two dimensional coupling can arise if the integrand of Eq. (1) is independent of $k_z$. Since the polarization of the LA and TA phonon modes is in-plane, the deformation coupling is to in-plane momentum,  $g_k=Dk\rightarrow D_{\perp}k_{\perp}$. In addition, for phonon energies $\hbar\omega_k > 10~\mathrm{meV}$ the dispersion curves of the LA and TA-phonons are also independent of $k_z$ \cite{Serrano_prl}. Hence, for most energies of interest the electron-phonon coupling is effectively two-dimensional. 

If we assume that only LA-phonons contribute, we extract a value $D_{2D}\approx 10.5~\mathrm{eV}$, using $\rho_{2D}=0.76~\mathrm{mg.m^{-2}}$ \cite{footnote_massdensity}. This is reasonable, since deformation coupling is usually much stronger for LA than TA-phonons. $\sigma_{LA}=0.35~\mathrm{nm}$ which is about 2.4 times the nearest neighbor separation of 0.144 nm \cite{Geick_pr}.

The deformation coupling strength is close to the {\it D=11 eV} found for UV-emitting defect-bound excitons coupled to 3D phonon-bath in bulk h-BN \cite{Vuong_prl}, and is close to values found for excitons in GaAs QDs \cite{Ramsay_prl}. However, it is relatively large compared to deformation coupling constants of up to 0.6 eV measured under static strain conditions for hBN color centers emitting at 2.14 eV \cite{Grosso_ncomm}. We also note that the deformation constant $D$ needed to reproduce the data increases with dimension, implying that the electron-phonon interaction is enhanced for lower dimension systems.


\section{Analysis of optical phonon sideband}

To gain insight into the interaction of the color center with optical phonons, we compare the dispersion curve of hBN measured in ref. \cite{Serrano_prl} to a close-up of the optical phonon sideband of the emission spectra taken at 20 K, see fig. 3. The red and green curves indicate bulk LO-phonon branches with $E_{2g}$ and $E_{1u}$ symmetry at the $\Gamma$-point,  where neighboring planes oscillate in-phase  or anti-phase, respectively. The black curve shows the bulk TO-branch. The over-bending of the in-phase LO($E_{2g}$)-band results from a spring-constant that changes sign with the separation between lattice sites, see table I of ref. \cite{Michel_prb2009}. For multi-layers \cite{Michel_prb2011}, adjacent out-of-phase layers ($k_z \neq 0$) generate an in-plane electric-field via the Coulomb interaction increasing the spring constants. This increases the energy of the $LO(E_{1u})$ with respect to the $LO(E_{2g})$ band, suppresses the over-bending, and results in energies that depend on the number of layers, as seen for example in other 2D materials such as graphene \cite{Ferrari_prl}, $MoS_2$, and $WS_2$ \cite{Sanchez_prb2011,Lee_ACSnano}.
 Several peaks can be observed. To aid identification, construction lines are drawn from the turning points in the bulk dispersion curves, corresponding to energies where the phonon density of states is high. 


\begin{figure}[ht!]
\begin{center}
\vspace{0.2 cm}
\includegraphics[width=1\columnwidth]{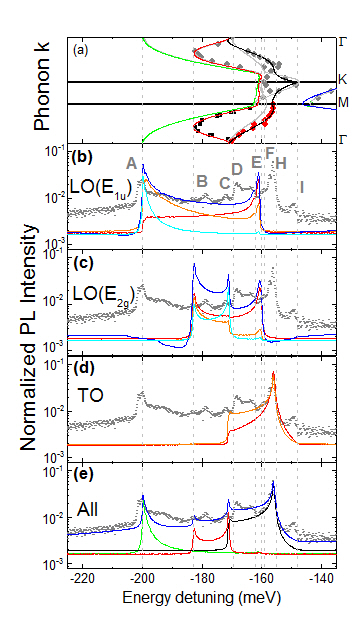}
\vspace{0.2 cm}
\end{center}
\caption{(a) Dispersion curves of the optical phonon branches. Lines are drawn to indicate turning points where the density of states is high. Comparison of data to calculations considering a single band: (b) $LO(E_{1u})$ (c) $LO(E_{2g})$ (d) TO. We consider a Fr\"{o}hlich (blue,cyan) and deformation (red,orange) coupling with (blue, $\epsilon_{eff}=22.2,\sigma=0$); (cyan,$\epsilon_{eff}=33.3, \sigma=0.2~\mathrm{nm}$);(red, $M=72~\mathrm{eV.nm^{-1}}$); (orange,$M=341.5~\mathrm{eV.nm^{-1}}$). For the TO-branch only the deformation coupling is presented. (e) A fit to data assuming Fr\"{o}hlich coupling to two LO-branches, and deformation coupling to the TO. Values of $\epsilon_{eff}(LO(E_{1u}))=33.3, \epsilon_{eff}(LO(E_{2g}))=66.7, \sigma_{LO}=0.2~\mathrm{nm}$, $M_{TO}=153~\mathrm{eV.nm^{-1}}, \sigma_{TO}=0.1~\mathrm{nm}$ are used.
}\label{fig:fig3}
\end{figure}

\begin{center}
\begin{table}
\begin{tabular}{|c|c|c|}
  \hline
  Peak & detuning (meV) & identity    \\
  \hline
  A & 200 & $\mathrm{LO}(E_{1u},\Gamma)$ \\
  B & 183 & $\mathrm{LO}(E_{2g},T)$  \\
  C & 178 & $ \mathrm{LO}(E_{2g},T)$ \\
  D & 172 & $\mathrm{LO}(E_{1u},\Gamma)/ \mathrm{TO}(\Gamma)$ ? \\
  E & 169 & $\mathrm{LO}(E_{1u},\Gamma)/ \mathrm{TO}(\Gamma) $ Anti-cross? \\
  F & 163 & $\mathrm{LO(M,K)}$ \\
  G & 159 & $\mathrm{LO(E_{1u},K)}$  \\
  H & 156 & TO(M),TO(AC)   \\
  I & 149 & $\mathrm{TO}(K),\mathrm{LO}(E_{2g},K)$ \\
  \hline
\end{tabular}
\caption{List of optical phonon peaks observed in fig. 3. All peaks can be linked to turning points in LO-branches, or where branch has strong LO-component. }
\end{table}
\end{center}

Table I compares the features labelled A-I with turning points in the dispersion curves \cite{Serrano_prl} of the optical sidebands, where the phonon density of states are high. The strong peak-A at 200 meV, also seen in \cite{Feldman_ArXiv} can be unambiguously identified as the $LO(E_{1u})$ band at the $\Gamma$-point of bulk h-BN. If h-BN behaves similarly to other  2D materials,  then this suggests that the defect resides in a multi-layer with $\mathcal{N}\geq 4$ \cite{Ferrari_prl,Sanchez_prb2011,Lee_ACSnano}. 

The lowest energy peak (I) corresponds to TO(K). The strongest peak (H) at 156 meV correspond to the TO-branch near to the Brillouin edge.  The relative strength of this peak compared to the LO-peak (A) is suggestive of a sample with few layers, rather than a single layer, as the relative DOS of the TO-branch compared with the LO-branches is higher since the degeneracy of the TO-branches is not lifted by the Coulomb interaction \cite{Michel_prb2011}. The presence of peak (E) suggests the turn-on of the electron-TO interaction at low momentum as the phonon energy decreases.

To model the electron-phonon interaction, we consider a Fr\"{o}hlich-like coupling to the LO-branch of dimension $d=2$ given by \cite{Peeters_prb}
\begin{eqnarray}
\vert g_k\vert^2=\frac{e^2\omega}{2^{4-d}\hbar\epsilon_0 V_d k^{d-1}\epsilon_{eff}}f(k) \label{eq:frohlich}
\end{eqnarray}
where $\epsilon_{eff}$ is treated as a fitting parameter. For a single conduction band electron
\begin{equation}
\frac{1}{\epsilon_{eff}}=\frac{1}{\epsilon_{\infty}}-\frac{1}{\epsilon_{s}}
\end{equation}
where $\epsilon_{\infty}$ and $\epsilon_s$ are the high frequency and static dielectric constants respectively. The other parameters are  the permittivity of free-space $\epsilon_0$, and the electron charge $e$. Note that qualitatively, there is no difference in $\gamma(t)$ for a 2D or 3D Fr\"{o}hlich interaction, and we cannot use the data to distinguish between them. However, the polarization of the LO-branch is in-plane, hence the argument in Eq. (\ref{eq:frohlich}) is $k\rightarrow k_{\perp}$. To model multiple layers, $\omega(k)\rightarrow \omega_{\alpha}(k_{\perp})$, where $\alpha$ indexes the $\mathcal{N}$ LO-branches according to the relative phase between layers, (i.e. $k_z$), $V_d\rightarrow \mathcal{N}V_2$, and a sum over $\mathcal{N}$ LO-branches is made in Eq. (1).

In addition to the Fr\"{o}hlich interaction, it is expected that both the LO and TO modes will exhibit a coupling proportional to the optical displacement. For optical-phonons, this is often referred to as deformation coupling \cite{Zhang_jpcm}, but can be treated as the $\mathcal{M}$-term in Eq. (2).

Figure 3 compares calculations of the different bands contributions to the spectrum. The contribution to $\gamma(t)$ of both the Fr\"{o}hlich interaction and the optical deformation coupling depends mostly on the inverse group velocity of the band. In fig. 3(d), the TO-branch is unaffected by additional layers, so should be less open to interpretation. The strongest peak (H) is reproduced with  $\mathcal{M}$=72-341 $\mathrm{eV.nm^{-1}}$, depending on the $\sigma_{TO}$ used. This appears high, however the origin of the coupling is the deformation of the lattice, and at the Brillouin edge the `acoustic' deformation coupling required to give the same shift in energy is: $D=Ma/\pi= 3.3-15.6 ~\mathrm{eV}$, which is not unreasonable.

In fig. 3(b), it is clear that 200-meV peak can only be explained by Fr\"{o}hlich interaction to a bulk-like $LO(E_{1u})$ mode, since the M-term has no peak at the $\Gamma$-point where $k_{\perp}=0$.    Naively, one would expect the defect to couple to all of the LO-modes equally. In fig. 3(c), the $E_{2g}$ band has a higher DOS since the spread in energies is lower. Hence if the coupling were equal, this would imply that the peak at the LO(T) point would be stronger than the 200-meV peak. This implies that for this defect, the  Fr\"{o}hlich coupling is stronger for out-of-phase layers. This may suggest that dipole fields generated by adjacent layers cancel or enhance the E-field generated by LO-phonon. Or, it may relate to which layer the defect resides. The strength of the interaction given by $\epsilon_{eff}^{-1}=0.03$ is small compared with the value of 0.056 for a single electron in bulk h-BN \cite{Ioffe}. This is to be expected since it is the difference in the charge distribution of the energy-levels of optical transition that matters.

Figure 3(e) presents a fit to the optical sideband using two LO bands with Fr\"{o}hlich coupling and a TO-band with deformation coupling. We note that the model is over-specified, and the numbers used should not be considered as accurate. The discrepancy at high energy may suggest that more layers need to be considered, or that $\sigma_{LO}<0.2~\mathrm{nm}$.



\section{Emitter II}

\begin{figure}[ht!]
\begin{center}
\vspace{0.2 cm}
\includegraphics[width=1.0\columnwidth]{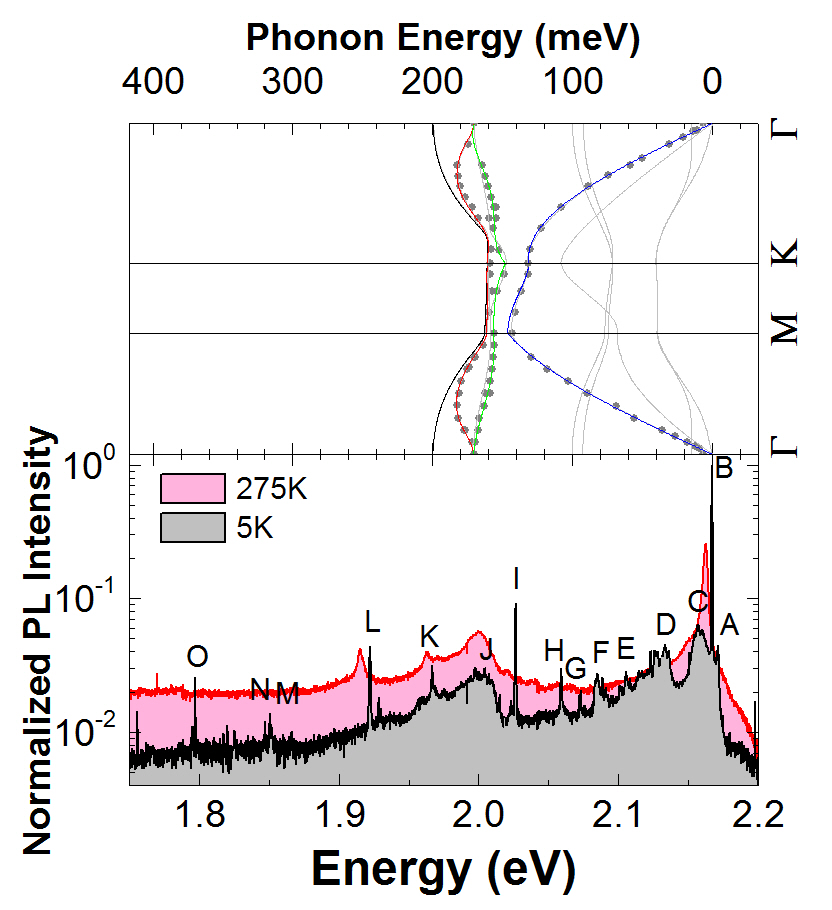}
\vspace{0.2 cm}
\end{center}
\caption{   Photoluminescence spectra of emitter II. The peaks are identified in table II. }\label{fig:figEmB}
\end{figure}

\begin{center}
\begin{table}
\begin{tabular}{|c|c|c|}
  \hline
  Peak & detuning (meV) & identity    \\
  \hline
  A & -5.5 & ZPL$_z$ \\
  B & 0 & ZPL$_x$  \\
  C & 9.8 & $\bar{ZA}(\Gamma)_z$ or $ZA(A)_x$ \\
  D & 32.7 & $ZA(M,K)_z$ \\
  E & 61 & $TA(M)_z$?\\
  F & 83.2 & ? \\
  G & 94 & $\bar{ZO}(\Gamma)_z$ or $ZO(\Gamma)_x$  \\
  H & 107.6 & $TA(K)_x$   \\
  I & 140 & $LA(M)_x$ \\
  J & 159 & $TO(M)_x$ \\
  K & 200 & $\bar{LO}(\Gamma)$\\
  L & 244.5 & $LA(M)_x+TA(K)_x$ \\
  M & 315.5 & 2BZ Opt\\
  N & 319.8 & $TO(K)_x+[LO(\Gamma),TO(\Gamma)]$ or 2LO(T)\\
  O & 369.4 & $LO(\Gamma)+TO(\Gamma)$\\
  \hline
\end{tabular}
\caption{List of peaks identified in fig. 4 for emitter II. Peaks $A$ and $B$ are identified with ZPL lines labeled $z$ and $x$. The other peaks are identified relative to these two peaks. The bar labels modes where layers oscillate in anti-phase. }
\end{table}
\end{center}

Figure 4 presents the photoluminescence spectrum of emitter II. Although the emission energies are similar, the spectrum is more complicated than for emitter-I.
If we assign peaks A and B to ZPLs, then peaks C-F could be explained as out-of-plane polarized ZA  and ZO phonon modes coupled to ZPL-A. Peaks H to O can be explained in terms of in-plane polarized phonons coupled to peak-B.
This suggest that peak-A is an out-of-plane polarized transition, which is weaker than the in-plane polarized peak-B, with a separation of 5.5 meV. For an out-of-plane transition, one of the energy-levels has a $p_z$ orbital \cite{Abdi_ArXiv}, resulting in one energy-level that is more sensitive to out-of-plane lattice displacements, and hence the emission energy couples to z-polarized phonons. Likewise for the in-plane polarized dipole. We note that peak (I) $LA(M)$ and peak (H) $TA(K)$ and peak (L) $LA(M)+TA(K)$ are especially strong, and correspond to displacements along the bond direction of the heavier Nitrogen sub-lattice. Observation of the LA(M)-peak suggests that the change in the charge distribution is aligned along the bond direction.  Once again, there is a 200-meV peak of the $\bar{LO}(\Gamma)$ point, indicating the sample has  a few layers. Otherwise the in-plane optical phonon sideband has fewer features than emitter I, possibly this is due to broadening with ZA-phonons, or the emitter preferentially couples to TO along the K-direction where the DOS is less peaked.

This suggests that either emitter I and II are from different species of defect, despite their similar emission energies 2.171 vs (2.167 or 2.171) eV. Or the local environment of emitter II somehow activates the out-of-plane transition.





\section{Conclusions}

The phonon sidebands of two color centers in h-BN emitting close to 2.17 eV have been analyzed.
For emitter I, the acoustic sideband can be described by deformation coupling to an effective two-dimensional phonon-bath with exponential form-factor.
This arises because the phonons are polarized in-plane, and for phonon energies larger than 10 meV the acoustic phonon dispersion is degenerate with respect to out-of-plane momentum.  For a two-dimensional system, this  results in an intrinsic sub-Lorentzian broadening of the ZPL \cite{Palma_prsoc}, and could limit the optimum photon indistinguishability that could be achieved. However, looking at the dispersion curves, one might expect a crossover to an effective 3D phonon bath at low phonon energies. There may be implications for the interpretation of the $\gamma_{ZPL}\sim T^3$ temperature dependence that has been reported elsewhere \cite{Jungwirth_nl,Sontheimer_prb}, and is confirmed in our studies. In refs. \cite{Sontheimer_prb}, by comparison with defects in diamond \cite{Neu_njp,Muller_prb,Jahnke_njp}, this behavior is attributed to a defect moving in an E-field generated by nearby charges undergoing a 2-phonon transition facilitated by an auxiliary state of energy $\Delta$ with respect to one of the energy-levels of the optical transition \cite{Jahnke_njp}. If this is the case, the power-law will depend on the dimensionality of the phonon bath, and this interpretation may not hold.

For emitter II, there appears to be two ZPL separated by 5.5 meV. One couples to in-plane phonons, and the other to out-of-plane. We cautiously suggest that this is related to the orientation of the optical dipole.

The LO-band exhibit a number of peaks that can be identified with turning points in the dispersion curve. With further work to analyze how different defects couple to different phonon modes, the phonon modes expressed in the PL spectra may help to identify the defect. A peak at 200meV, which can be attributed to Fr\"{o}hlich coupling to an LO-phonon where neighboring layers oscillate in anti-phase is prominent in both emitters, and is reported in ref. \cite{Feldman_ArXiv}. This may be a signature of a multi-layer flake with $N>3$. The LO-phonons are Fr\"{o}hlich coupled, and TO-phonons can be described by a deformation coupling proportional to the lattice displacement.

In ref. \cite{Feng_nl}, in monolayer hBN samples where the emitters are at approximately 580nm, the dominant defects in TEM was found to be the boron vacancy. Therefore, the defects studied here may be the $V_B^-$.

Finally, we note that the cut-off length is small $\sigma_{LA}\approx 0.4~\mathrm{nm}$, and indicates small carrier wavefunctions. Due to the high speed-of-sound, this corresponds to a cut-off temperature of $\hbar\omega_c/k_B\approx 400 K$, explaining the temperature robustness of the ZPL.

\acknowledgements
We gratefully acknowledge financial support from the Engineering and Physical Sciences Research Council UK under grants EP/S001557/1 and EP/P026656/1. We thank J. A. Haigh for critical reading of the manuscript.


\begin{thebibliography}{1}
\bibitem{footnote_massdensity}
The mass density is estimated using a nearest neighbor lattice constant of 0.144nm \cite{Geick_pr}
\end{thebibliography}
\end{document}